\documentclass[journal,comsoc,onecolumn]{IEEEtran}
\usepackage[T1]{fontenc}

\usepackage[cmintegrals]{newtxmath}

\usepackage{amsmath,amssymb}

\usepackage{bbm,bm}
\usepackage{pgfplots}
\usepackage{siunitx}
\usepackage{graphicx}
\usepackage{algorithm,algpseudocode}
\usepackage{mathtools}
\usepackage{url}
\usepackage[export]{adjustbox}
\usepackage{color}

\ifCLASSOPTIONcompsoc
 \usepackage[caption=false,font=normalsize,labelfont=sf,textfont=sf]{subfig}
\else
 \usepackage[caption=false,font=footnotesize]{subfig}
\fi

\usepackage{tikz,tkz-graph}
\usetikzlibrary{matrix,positioning,fit,calc,shapes}
\pgfplotsset{compat=newest}
\let\oldbrace\{
\def\{{\oldbrace\kern0.5pt}

\def\Cov{\mathop{\rm Cov}\nolimits}%
\newtheorem{example}{Example} 
\newcommand{\ve}{\bm}
\newtheorem{theorem}{Theorem}
\newtheorem{remark}{Remark}
\newtheorem{lemma}{Lemma}
\newtheorem{proposition}{Proposition}
\newtheorem{corollary}{Corollary}
\newtheorem{definition}{Definition}

\begin{document}

\title{Sum-Rate Capacity for Symmetric Gaussian Multiple Access Channels with Feedback}

\author{Erixhen~Sula,~\IEEEmembership{Student Member,~IEEE,}
	     Michael~Gastpar,~\IEEEmembership{Fellow,~IEEE,}
           ~Gerhard~Kramer,~\IEEEmembership{Fellow,~IEEE}
\thanks{Erixhen Sula is with \'Ecole Polytechnique F\'ed\'erale de Lausanne (EPFL), Switzerland,
e-mail: erixhen.sula@epfl.ch.}
\thanks{Michael Gastpar is with \'Ecole Polytechnique F\'ed\'erale de Lausanne (EPFL), Switzerland,
e-mail: michael.gastpar@epfl.ch}
\thanks{Gerhard Kramer is with Technical University of Munich (TUM), Germany,
e-mail: gerhard.kramer@tum.de}
}

\markboth{IEEE TRANSACTIONS ON INFORMATION THEORY,~Vol.~00, No.~0, Month~2018}%
{Author \MakeLowercase{\textit{et al.}}: Sum-Rate Capacity for Gaussian Multiple Access Channel with Feedback}

\maketitle

\begin{abstract}
The feedback sum-rate capacity is established for the symmetric $J$-user Gaussian multiple-access channel (GMAC). The main contribution is a converse bound that combines the dependence-balance argument of Hekstra and Willems (1989) with a variant of the factorization of a convex envelope of Geng and Nair (2014). The converse bound matches the achievable sum-rate of the Fourier-Modulated Estimate Correction strategy of Kramer (2002).
\end{abstract}

\IEEEpeerreviewmaketitle

\section{Introduction}
\IEEEPARstart{T}{he} feedback capacity of the \emph{two-user} Gaussian multiple-access channel (GMAC) was established by Ozarow~\cite{Ozarow}. The coding theorem was based on extending feedback strategies of Elias~\cite{Elias} and Schalkwijk and Kailath~\cite{Schalkwijk--Kailath} (see~\cite{Gallager--Nakiboglu}), while the converse followed from a cut-set argument. Kramer~\cite{Kramer} extended Ozarow's scheme to more than two users by using a method he called Fourier-Modulated Estimate Correction, or Fourier-MEC. For the symmetric GMAC and sufficiently large signal-to-noise ratio (SNR), the Fourier-MEC sum-rate meets the cut-set bound and is thus optimal. However, the problem remains open for low and intermediate SNRs.

The coding schemes in~\cite{Ozarow,Elias,Schalkwijk--Kailath,Kramer} start by mapping the message onto a point on the real line or complex plane, and they iteratively correct the receiver's estimate of this point by using linear minimum mean squared error (LMMSE) estimation. There are many variants of the schemes. For example, one can convert complex-channel strategies to real-channel strategies~\cite{Kramer}, one can interpret the LMMSE step as posterior matching \cite{Shayevitz--Feder,Lan}, and one can use multi-dimensional Fourier transforms for Fourier-MEC, e.g., a Hadamard transform~\cite{Kramer,Lan}.

Progress on improving capacity outer bounds was made in~\cite{Kramer--Michael2006} by applying the Hekstra-Willems dependence-balance argument~\cite{Hekstra--Willems}, and in~\cite{Wigger--Kim} that studies linear feedback strategies. Other results regarding GMACs with imperfect and/or noisy feedback are presented in~\cite{Lapidoth--Wigger,Tandon--Ulukus2009,Tandon--Ulukus2011}. Inner and outer bounds for other classes of multiple-access channels (MACs) with feedback are derived in~\cite{Cover--Leung,Kramer2003,Tandon--Ulukus}. 

\subsection{Contribution}
We derive a converse bound that establishes the following capacity result. Achievability follows from \cite[Sec.~V]{Kramer}.
\begin{theorem} \label{thm-main}
The feedback sum-rate capacity of the $J$-user symmetric GMAC is
\begin{align}
C_\text{sum} = \frac{1}{2} \log_2\left(1+PJ \beta\right) \quad \text{bits/channel use}
\end{align}
where $\beta$ is the unique solution satisfying $\beta\in[1,J]$ and
\begin{align} \label{eq:imp}
\left(1+PJ\beta \right)^{J-1}=(1+P \beta (J-\beta))^J.
\end{align}
\end{theorem}
The proof of Theorem~\ref{thm-main} combines the Lagrange-duality approach of~\cite{Gastpar--Kramer} with a variant of the factorization of a convex envelope used in~\cite{Geng--Nair,Courtade}, and that was inspired by work on functional inequalities \cite{Lieb,Carlen}. One difference to~\cite{Geng--Nair} is that in our problem the matrix that describes the channel is rank-deficit. Besides establishing the optimality of Gaussian signaling, we transform a non-convex Lagrangian dual problem into a convex problem.

\subsection{Notation}
We use the following notation. Random variables are denoted by uppercase letters and their realizations by lowercase letters. Random column vectors are denoted by boldface uppercase letters and their realizations by boldface lowercase letters. The $i$-th entry of the column vector $\ve X$ is denoted by $\ve X_i$ or $[\ve X]_i$. We denote matrices with uppercase letters, e.g., $A,B,C$. The $(i,j)$ element of matrix $A$ is denoted by $A_{ij}$ or $[A]_{ij}$. For the cross-covariance matrix of two random vectors $\ve X$ and $\ve Y$, we use the shorthand notation $K_{\ve X \ve Y}$, and for the covariance matrix of a random vector $\ve X$ we use the shorthand notation $K_{\ve X}:= K_{\ve X \ve X}$. Calligraphic letters denote sets, e.g., $\mathcal{A},\mathcal{B},\mathcal{C}$. The expression $X \sim \mathcal{N}(m,\sigma^2)$ denotes a Gaussian random variable with mean $m$ and variance $\sigma^2$. We denote the convergence in distribution (weak convergence) by $\overset{w}{\Rightarrow}$. 

\subsection{Organization}
This paper is organized as follows. In Section \ref{sec:2}, we present the system model and review existing capacity bounds. In Section \ref{sec:4}, we give an upper bound on the sum rate for general GMACs with feedback. In Section \ref{sec:sym}, we prove Theorem~\ref{thm-main}. Section \ref{sec:con} concludes the paper and the appendices provide supporting results and proofs.

\section{System Model and Related Work} \label{sec:2}
Consider a GMAC with $J$ transmitters (called users) with channel input symbols $X_1,X_2,\dots,X_J$, and a receiver with the channel output symbol $Y$. The received signal at time instant $i$ is 
\begin{equation}
Y_i = Z_i + \sum_{j=1}^{J} g_j X_{j,i}
\label{eq:channel}
\end{equation} 
for $i=1,2,\dots,n$, where $Z_1,Z_2,\dots,Z_n$ is a string of independent and identically distributed (i.i.d.) zero-mean Gaussian noise variables with unit variance and $g_1,g_2,\dots,g_J$ are channel gains. The $J$ channel inputs have the block power constraints
\begin{equation}
\sum_{i=1}^{n} E\left[X_{j,i}^2\right] \leq nP_j,  \quad {j=1,2,\dots,J}.
\label{eqn:pc}
\end{equation}
The SNR of user $j$ is thus $P_j g_j^2$.
If $P_1 g_1^2 = P_2 g_2^2 = \ldots = P_J g_J^2$, then the transmitters can be swapped without changing the problem. For such models, we may as well set $P_j=P$ and $g_j=1$ for all $j$, and we refer to this channel as the \emph{symmetric} GMAC.

Let $W_j$ with $nR_j$ bits be the message of user $j$. The transmitted signal at time instant $i$ is
\begin{equation}
X_{j,i}=f_{j,i}(W_j, Y^{i-1}), \quad {j=1,2,\dots,J}
\end{equation}  
where the $f_{j,i}(\cdot)$ are encoding functions to be optimized. The receiver puts out the estimates
\begin{equation}
   \left( \hat{W}_1,\hat{W}_2,\ldots,\hat{W}_J \right) = g(Y^n)
\end{equation}
where $g(\cdot)$ is a decoding function.
The event that the receiver makes an error is
\begin{equation}
   \mathcal{E} = \bigcup_{j=1}^J \left\{\hat{W}_j \ne W_j \right\}.
\end{equation}
The rate-tuple $\ve R = (R_1,R_2,\ldots,R_J)$ is said to be \emph{achievable} if, for any specified positive error probability $P_e$ and sufficiently large $n$, there are encoding functions and a decoder such that $\text{Pr}\left[\mathcal{E}\right] \le P_e$. The closure of the set of achievable $\ve R$ is called the \emph{capacity region} $\mathcal{C}_\text{MAC-FB}$. We are interested in characterizing the \emph{sum-rate capacity} $C_\text{sum}$, i.e., the maximum sum of the entries of any $\ve R$ in $\mathcal{C}_\text{MAC-FB}$.

%

\subsection{Capacity without Feedback}
The capacity region $\mathcal{C}_\text{MAC}$ \emph{without} feedback is the set of $\ve R = (R_1,R_2,\dots,R_J)$ satisfying (see \cite[Section~14.3]{Cover--Thomas})
\begin{equation}
R_{\mathcal{S}} \leq \frac{1}{2} \log(1+P_{\mathcal{S}})
\end{equation}
for all $\mathcal{S} \subseteq \mathcal{J} = \left\{1,2 ,\dots,J \right\}$, where $R_{\mathcal{S}}= \sum_{j \in \mathcal{S}} R_j$ and $P_{\mathcal{S}}= \sum_{j \in \mathcal{S}} P_j$. The sum-rate capacity without feedback is therefore
\begin{equation}
\frac{1}{2} \log(1+P_{\mathcal{J}}).
\end{equation}

\subsection{Two-User Capacity with Feedback}
Feedback allows the users cooperate to increase rates. For $J=2$ the capacity region is known to be \cite{Ozarow}
\begin{align} \label{eqn:cf-j2}
\mathcal{C}_\text{MAC-FB}
& = \bigcup_{0 \le \rho \le 1} \mathcal{R}(\rho)
\end{align}
where $\mathcal{R}(\rho)$ is the set of rate pairs $(R_1,R_2)$ that satisfy
\begin{equation}
\label{eqn:MAC-J2-regionG}
\begin{split}
& 0 \le R_1 \le \frac{1}{2} \log\left(1+P_1(1-\rho^2) \right) \\ 
& 0 \le R_2 \le \frac{1}{2} \log\left(1+P_2(1-\rho^2) \right) \\
& R_1+R_2 \le \frac{1}{2} \log\left(1+P_1+P_2 +2\rho \sqrt{P_1P_2} \right).
\end{split}
\end{equation}
The parameter $\rho$ is the correlation coefficient of $X_1$ and $X_2$, and the optimal $X_1$ and $X_2$ are zero-mean Gaussian with second moments $P_1$ and $P_2$, respectively. $\mathcal{C}_\text{MAC-FB}$ is here the same as a standard cut-set bound. However, we show that cut-set bounds are loose for $J>2$, and that dependence balance bounds can characterize the fundamental limits of communication.

\subsection{Two-User Dependence Balance Bounds} 
Dependence balance bounds were introduced by Hekstra and Willems~\cite{Hekstra--Willems} for single output two-way channels. The tool generalizes to other models such as MACs with feedback. For example, for the two-user MAC with feedback, the achievable $(R_1,R_2)$ must satisfy
\begin{equation} \label{eqn:MAC-J2-region}
\begin{split} 
& 0 \le R_1 \le I(X_1;Y|X_2,T)  \\
& 0 \le R_2 \le I(X_2;Y|X_1,T)  \\
& R_1+R_2 \le I(X_1,X_2;Y|T) 
\end{split}
\end{equation}
for some $p(t,x_1,x_2,y)$ for which 
\begin{align}
& T \leftrightarrow [X_1,X_2] \leftrightarrow Y \text{ forms a Markov chain} \\
& I(X_1;X_2 | T) \le I(X_1;X_2|Y,T). \label{eqn:con}
\end{align}
In \cite[Section~7]{Hekstra--Willems}, the term $I(X_1;X_2 | T)$ is interpreted as the amount of dependence \emph{consumed}, and $I(X_1;X_2 | Y, T)$ as the amount of dependence \emph{produced} by communication. An interpretation of the inequality \eqref{eqn:con} is thus that dependence consumed cannot exceed the dependence produced, i.e., communication is limited by \emph{dependence balance}. 

Other interpretations of this bound are described in \cite{Kramer--Michael2006}. Observe that (\ref{eqn:con}) can be rewritten in the following two ways:
\begin{align}
& I(X_1;Y|T) +I(X_2;Y|T) \le I(X_1,X_2;Y|T)
\label{eqn:resh0} \\
& I(X_1,X_2;Y|T) \le I(X_1;Y|X_2,T) +I(X_2;Y|X_1,T). \label{eqn:resh}
\end{align}
The bound \eqref{eqn:resh} requires the set function $f: 2^{\{1,2\}} \rightarrow \mathbb{R}$ defined by
\begin{align}
f(\{1\}) & = I(X_1;Y|X_2,T) \\
f(\{2\}) & = I(X_2;Y|X_1,T) \\
f(\{1,2\}) & = I(X_1,X_2;Y|T) 
\label{eqn:submodular}
\end{align}
to be \emph{submodular}. In other words, for valid choices of $p(t,x_1,x_2,y)$, the rate region defined by \eqref{eqn:MAC-J2-region} is a \emph{polymatroid}. We may thus interpret dependence balance as a submodularity (or polymatroid) constraint, i.e., communication is limited by submodularity.

We remark that for two-user GMACs the dependence balance bound yields the same rate region as the standard cut-set bound. However, the dependence balance bound is more informative in the following sense. Consider jointly Gaussian $p(t,x_1,x_2,y)$. The optimal correlation coefficient $\rho^*$ in \eqref{eqn:MAC-J2-regionG} is the one that satisfies \eqref{eqn:con}-\eqref{eqn:resh} with equality. However, dependence balance (or submodularity) limits $\rho$ to the range $[0,\rho^*]$, whereas the cut-set bound permits all $\rho$ in $[0,1]$.

\subsection{Multi-User Dependence Balance Bounds} \label{sec:main-to-rely}
The two-user dependence balance concept was generalized to $J$ users in \cite[Thm.~4]{Kramer2003} and more dependence balance bounds are derived in \cite[Thm.~1]{Kramer--Michael2006}. The capacity region of the $J$-user MAC with feedback is a subset of the set of rate-tuples $(R_1,R_2,\dots,R_J)$ satisfying 
\begin{align}
R_{\mathcal{S}} \leq I(X_{\mathcal{S}};Y|X_{\mathcal{S}^C},T)
\end{align}
for all $\mathcal{S} \subseteq \mathcal{J}$, where $\mathcal{S}^C$ is the complement of $\mathcal{S}$, and where 
\begin{align}
& T \leftrightarrow [X_1,X_2,\dots,X_J] \leftrightarrow Y \text{ forms a Markov chain} \\
& I(X_1,X_2,\dots,X_J;Y | T) \leq \frac{1}{M-1} \sum_{m=1}^M I(X_{\mathcal{S}_m^C} ; Y|X_{\mathcal{S}_m},T), \label{eqn:bes}
\end{align}
for any partition $\left\{ \mathcal{S}_m \right\}_{m=1}^M$ of $\mathcal{J}$ into $M\geq 2$ subsets. One may again interpret \eqref{eqn:bes} as a submodular constraint.
For example, for the partition $\mathcal{S}_1=\left\{ 1 \right\}$, $\mathcal{S}_2=\left\{ 2 \right\}$, $\dots$, $\mathcal{S}_J=\left\{ J \right\}$ the dependence balance constraint \eqref{eqn:bes} becomes
\begin{align}
I(X_1,X_2, \dots, X_J;Y|T) \leq \frac{1}{J-1} \sum_{j=1}^J I(X_{\mathcal{J}\backslash  \left\{j\right\} };Y|X_j,T)
\end{align}
where $\mathcal{J}\backslash  \left\{j\right\}$ is the set $\left\{1,2, \dots,j-1,j+1,\dots,J \right\}$. As usual, one can add the power constraints $E[X_j^2]\le P_j$, $j\in\mathcal{J}$, to these bounds.
Also, as for \eqref{eqn:resh0} for the two-user case, the bound \eqref{eqn:bes} can be written as
\begin{align}
\sum_{m=1}^M I(X_{\mathcal{S}_m};Y|T) \leq I(X_1,X_2, \dots, X_J;Y|T).
\label{eqn:ndr}
\end{align}

\subsection{Cut-set Bound}
The cut-set bounds give the following result, see~\cite{Ozarow} and \cite[Theorem~15.10.1]{Cover--Thomas}.
\begin{proposition}
For the two-user symmetric GMAC with feedback, we have
\begin{align}
C_\text{sum} \le \max\limits_{-1 \le \rho \le 1} \min & \Bigg\{ \underbrace{\frac{1}{2} \log (1+2P(1+\rho)}_\text{$f_1(\rho)$},  \underbrace{ \log (1+P(1-\rho^2))}_\text{$f_2(\rho)$} \Bigg\}.
\label{eqn:cut-set-J2}
\end{align}
\end{proposition}
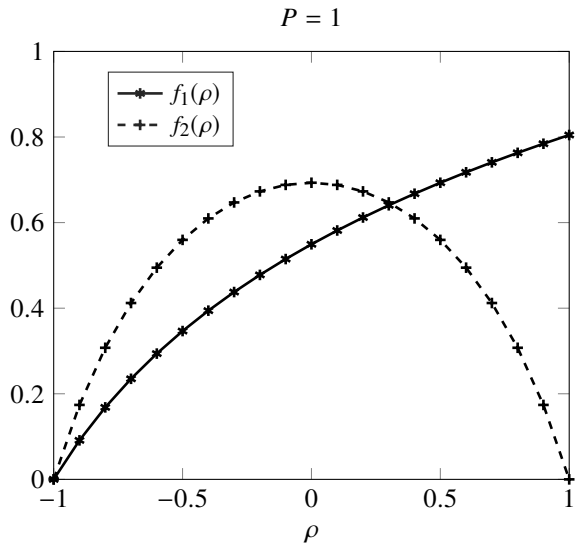
\begin{figure}[t]
\centering
{\begin{tikzpicture}
\begin{axis}[%
xmin=-1,
xmax=1,
xlabel={$\rho$},
ymode=linear,
ymin=0,
ymax=1,
title={$P=1$},
legend style={at={(0.105,0.7777)}, anchor=south west, legend cell align=left, align=left, draw=white!15!black, style={font=\small},}
]
\addplot [ line width=1.0pt,mark=asterisk]
  table[row sep=crcr]{%
-1	0\\
-0.9	0.0911607783969773\\
-0.8	0.168236118310606\\
-0.7	0.235001814622868\\
-0.6	0.29389333245106\\
-0.5	0.346573590279973\\
-0.4	0.394228680182135\\
-0.3	0.43773436867695\\
-0.2	0.477755722513718\\
-0.1	0.514809708590579\\
0	0.549306144334055\\
0.1	0.58157540490284\\
0.2	0.611887715811058\\
0.3	0.640466922731032\\
0.4	0.66750053336617\\
0.5	0.693147180559945\\
0.6	0.717542262644661\\
0.7	0.740802270462108\\
0.8	0.763028151747525\\
0.9	0.784307958956923\\
1	0.80471895621705\\
};
\addlegendentry{$f_1(\rho)$}

\addplot [ line width=1.0pt, dashed, mark=+,mark options={solid}] table[row sep=crcr]{%
-1	0\\
-0.9	0.173953307123438\\
-0.8	0.307484699747961\\
-0.7	0.412109650826833\\
-0.6	0.494696241836107\\
-0.5	0.559615787935423\\
-0.4	0.609765571620894\\
-0.3	0.647103242058539\\
-0.2	0.672944473242426\\
-0.1	0.688134638736401\\
0	0.693147180559945\\
0.1	0.688134638736401\\
0.2	0.672944473242426\\
0.3	0.647103242058539\\
0.4	0.609765571620894\\
0.5	0.559615787935423\\
0.6	0.494696241836107\\
0.7	0.412109650826833\\
0.8	0.307484699747961\\
0.9	0.173953307123438\\
1	0\\
};
\addlegendentry{$f_2(\rho)$}

\end{axis}
\end{tikzpicture}%
}
\caption{Cut-set bounds for the sum-rate of a two-user symmetric  GMAC with feedback.}\label{fig:man}
\end{figure}
The sum-rate on the RHS of \eqref{eqn:cut-set-J2} turns out to be achievable \cite{Ozarow}, and it is depicted in Figure \ref{fig:man}.
Similarly, again starting from~\cite[Theorem~15.10.1]{Cover--Thomas}, we obtain the following result.
\begin{proposition}\label{prop-cutset-3}
For the three-user symmetric GMAC with feedback, we have
\begin{align}
C_\text{sum} & \le \max\limits_{-1/2 \le \rho \le 1} \min \Bigg\{ \underbrace{\frac{1}{2} \log (1+3P(1+2\rho)}_\text{$g_1(\rho)$}, \nonumber \\
& \quad \underbrace{ \frac{3}{4} \log(1+2P(1- \rho )(1+2 \rho))}_\text{$g_2(\rho)$}, \underbrace{ \frac{3}{2} \log \left( 1 + \frac{P(1+2 \rho)( 1- \rho)}{1 + \rho}\right) }_\text{$g_3(\rho)$} \Bigg\}.
\label{eqn:three-user-UB}
\end{align}
\end{proposition}

\begin{figure}[t]
\centering
{
\begin{tikzpicture}
\begin{axis}[%
xmin=0,
xmax=0.2,
xlabel={$\rho$},
ymode=linear,
ymin=0.32,
ymax=0.42,
title={$P=0.3$},
legend style={at={(0.105,0.7477)}, anchor=south west, legend cell align=left, align=left, draw=white!15!black, style={font=\small},}
]
\addplot [ line width=1.0pt,mark=asterisk]
  table[row sep=crcr]{%
0	0.320926943086197\\
0.02	0.330311994427193\\
0.04	0.339524128090222\\
0.06	0.348569600914741\\
0.08	0.357454336170729\\
0.1	0.366183946856613\\
0.12	0.374763756998026\\
0.14	0.383198821149769\\
0.16	0.391493942279867\\
0.18	0.399653688194168\\
0.2	0.407682406642097\\
};
\addlegendentry{$g_1$}

\addplot [ line width=1.0pt, dashed, mark=+,mark options={solid}] table[row sep=crcr]{%
0	0.352502721934302\\
0.02	0.357883374745302\\
0.04	0.362781957226386\\
0.06	0.36720761916349\\
0.08	0.371168508565136\\
0.1	0.37467182361546\\
0.12	0.37772385766929\\
0.14	0.380330037897928\\
0.16	0.382494958096421\\
0.18	0.38422240607699\\
0.2	0.385515385996882\\
};
\addlegendentry{$g_2$}

\addplot [ line width=1.0pt, dashed, mark=-,mark options={solid}] table[row sep=crcr]{%
0	0.393546396701236\\
0.02	0.39327487891608\\
0.04	0.392480929626724\\
0.06	0.39119331833471\\
0.08	0.389438212003884\\
0.1	0.387239449778189\\
0.12	0.384618783000507\\
0.14	0.381596085546051\\
0.16	0.37818953866836\\
0.18	0.374415793887259\\
0.2	0.370290116897289\\
};
\addlegendentry{$g_3$}

\end{axis}
\end{tikzpicture}%
}\caption{Cut-set bounds for the sum-rate of a three-user symmetric GMAC with feedback.} \label{fig:man2}
\end{figure}
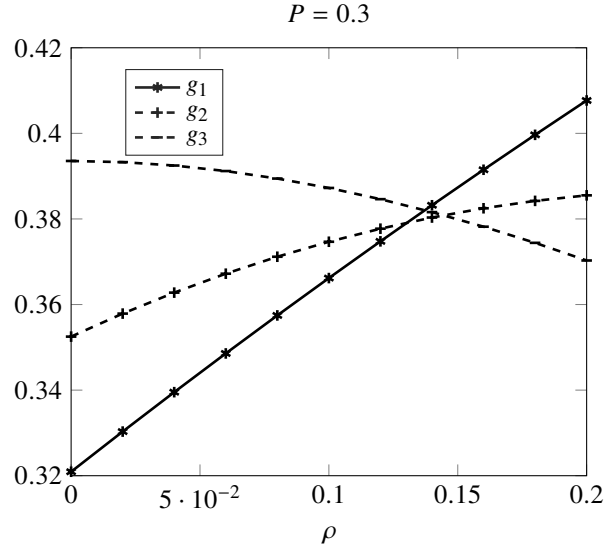

The sum rate \eqref{eqn:three-user-UB} is not generally achievable. Figure \ref{fig:man2} illustrates the situation for the special case $P=0.3.$ The cut-set bound of Proposition~\ref{prop-cutset-3} leads to an upper bound on the sum rate given by the intersection point of the curves $g_2$ and $g_3.$ However, we show that the capacity is given by the intersection of the curves $g_1$ and $g_2$; this intersection point is achieved by Fourier-MEC~\cite{Kramer}.

For the symmetric GMAC and large SNR, i.e., more that a certain threshold, the Fourier-MEC sum-rate meets the cut-set bound. For example, a sufficient condition for the cut-set bound to give the sum-rate capacity is that the SNR is greater than or equal to $2^{J+1}/J^2$~\cite{Kramer}. Observe that this threshold grows almost exponentially with the number of users.

\section{General Converse Bound for $J$ Users} \label{sec:4}
We derive the following upper bounds on the feedback sum-rate capacity of the general $J$-user GMAC.
\begin{theorem} \label{thm:gen}
For any $\lambda  \geq 0$ and any partition $\left\{ \mathcal{S}_m \right\}_{m=1}^M$ of $\mathcal{J}$ into $M\geq 2$ subsets, we have
\begin{align}
C_{\mathrm{sum}} \leq \max_{p(\ve{x}) \in G_{\mathcal{G}}} (1- \lambda)I(X_1,X_2,\dots,X_J;Y)+\frac{\lambda}{M-1}\sum_{m=1}^M I(X_{\mathcal{S}_m^C};Y|X_{\mathcal{S}_m})
\end{align} 
where $G_{\mathcal{G}}$ is the set of zero-mean Gaussian distributions satisfying $E[X_j^2] \leq P_j$ for $j=1,2,\dots,J$.
\end{theorem}

\subsection{Proof of Theorem \ref{thm:gen}} \label{subsec:bes}
Our converse bound starts from Section~\ref{sec:main-to-rely} and we use the shorthand $\ve X=(X_1,X_2,\ldots,X_J)$. We find it convenient to express our problem as a minimization, i.e., we seek to minimize $-I(\ve X;Y | T)$ over the input distributions $p(t,\ve x)$ that satisfy the dependence-balance constraint
\begin{align} \label{eqn:contgen2}
I(\ve X;Y|T) \leq \frac{1}{M-1} \sum_{m=1}^M I(X_{\mathcal{S}_m^C};Y|X_{\mathcal{S}_m},T)
\end{align}
for $(T,\ve X)$ such that $T \leftrightarrow \ve X \leftrightarrow Y$ forms a Markov chain. 
We will treat the power constraints in two steps. First, for any fixed covariance matrix $K$, we will optimize over all distributions satisfying $E[\ve X \ve X^T] = K$. Then, we will optimize over all $K$ whose diagonal entries are at most $P$.

Given the covariance matrix $K$, we form the Lagrangian for our optimization problem as
\begin{equation} \label{eq-def-lowerslambda}
\begin{split}
s_{\lambda}(\ve X|T):=& (\lambda -1) I(\ve X;Y|T)  -  \frac{\lambda}{M-1} \sum_{m=1}^M I(X_{\mathcal{S}_m^C};Y|X_{\mathcal{S}_m},T).
\end{split}
\end{equation} 
This can be rewritten as
\begin{align} 
s_{\lambda}(\ve X|T)=& -\left(\frac{\lambda}{M-1} +1\right) I(\ve X;Y|T)  +  \frac{\lambda}{M-1} \sum_{m=1}^M I(X_{\mathcal{S}_m};Y|T ). \label{eq-def-lowerslambda2}
\end{align}
We define
\begin{equation*}
S_{\lambda}(\ve X)=\inf\limits_{{\scriptstyle p{(t|\ve x)}:} \atop\scriptstyle T \leftrightarrow {\ve X} \leftrightarrow Y} \left\{ s_{\lambda}(\ve X|T) \right\}
\end{equation*}
and note that $S_{\lambda}(\ve X)$ is a convex function of $p(\ve x)$ because $S_{\lambda}(\ve X)$ is the lower convex envelope of $s_{\lambda}(\ve X)$, which is defined by dropping the random variable $T$ in \eqref{eq-def-lowerslambda}. In addition, we define 
\begin{align}
S_{\lambda}(\ve X|T)= \sum_{t} p(t) S_{\lambda}(\ve X| T=t).
\end{align}
The dual function of our problem for $K \succeq 0$ is
\begin{equation}
V_{\lambda}(K):=\inf\limits_{p{(\ve x)}:E[\ve X \ve X^T] = K } \left\{ S_{\lambda}(\ve X) \right\}.
\end{equation}
Alternatively, we have
\begin{equation} \label{eq-def-VlambdaK}
V_{\lambda}(K)=\inf\limits_{{\scriptstyle p{(t,\ve x)}:E[\ve X \ve X^T] = K} \atop\scriptstyle T \leftrightarrow {\ve X} \leftrightarrow Y} \left\{s_{\lambda}(\ve X|T)\right\}.
\end{equation}
By the standard Lagrangian duality we bound the original optimization problem as follows
\begin{align}
C_{\mathrm{sum}} &\leq - \max_{\lambda} \min_{\substack{p(t,\ve x):~E[X_j^2] \leq P_j \\ T {\leftrightarrow} {\ve X} {\leftrightarrow} Y }} s_{\lambda}(\ve X|T) \\
&=- \max_{\lambda} \min_{\substack{K \succeq 0: \\ [K]_{jj} \leq P_j}} \underbrace{\min_{\substack{p(t,\ve x):~E[\ve X \ve X^T] =K \\ T {\leftrightarrow} {\ve X} {\leftrightarrow} Y }} s_{\lambda}(\ve X|T)}_{V_{\lambda}(K)} .
\end{align}

From now on we mainly deal with the dual function $V_{\lambda}(K)$. For $0 < \lambda \le 1$, the minimization problem~\eqref{eq-def-VlambdaK} is a convex problem, and it follows from \eqref{eq-def-lowerslambda} and maximum entropy results that the optimizing distribution $p(t,x_1,x_2,x_3)$ is jointly Gaussian.
The more difficult case is $\lambda > 1$. Our approach for all cases will be to establish that the distribution attaining the minimum in~\eqref{eq-def-VlambdaK} must make the {\it channel input} Gaussian. This follows from a novel variant of the factorization of convex envelopes.

Consider two independent uses of the GMAC:
\begin{equation}
\begin{split}
Y_1&= G \ve X_1 + Z_1 \\
Y_2&= G \ve X_2 + Z_2
\end{split}
\end{equation}
where $G=\begin{bmatrix} 1 & 1 & \dots & 1 \end{bmatrix}$, $\ve X_1$ and $\ve X_2$ are independent and identically distributed, and where $Z_1,Z_2 \sim \mathcal{N}(0,1)$ are independent.
One key difference to~\cite{Geng--Nair} is that $G$ is \emph{not} an invertible matrix. We define
\begin{align}
\ve X_{\theta_1}&=\frac{1}{\sqrt{2}} (\ve X_1 + \ve X_2), \quad \ve X_{\theta_2}=\frac{1}{\sqrt{2}} (\ve X_1 - \ve X_2), \\
Y_{\theta_1}&=\frac{1}{\sqrt{2}} (Y_1 + Y_2), \quad Y_{\theta_2}=\frac{1}{\sqrt{2}} (Y_1 - Y_2).
\end{align}
We thus have
\begin{align}
Y_{\theta_1} = G \ve X_{\theta_1} + \tilde{Z}_1, \quad
Y_{\theta_2} = G \ve X_{\theta_2} + \tilde{Z}_2 \label{eq:Ytheta}
\end{align}
where $\tilde{Z}_1,\tilde{Z}_2 \sim \mathcal{N}(0,1)$ are independent. Moreover, we generalize the definition~\eqref{eq-def-lowerslambda2} to the two-letter extension as
\begin{align}
s_{\lambda}(\ve X_1, \ve X_2|T):=& -\left(\frac{\lambda}{M-1} +1\right) I(\ve X_1,\ve X_2;Y_1,Y_2|T) +  \frac{\lambda}{M-1} \sum_{m=1}^M I([ \ve X_1]_{\mathcal{S}_m},[ \ve X_2]_{\mathcal{S}_m};Y_1, Y_2|T ).
\end{align}
Next, we state and prove four key propositions.
\begin{proposition}
\label{propo:1}
$I(\ve X_1, \ve X_2; Y_1, Y_2)=I(\ve X_{\theta_1}, \ve X_{\theta_2};Y_{\theta_1}, Y_{\theta_2})$.
\end{proposition}
\emph{Proof}: The function $f(x,y)=\left( (x+y)/\sqrt{2}, (x-y)/\sqrt{2} \right)$ is bijective. $\hfill \square$

\begin{proposition} \label{prop:Mar}
The chain $Y_{\theta_1} {\leftrightarrow} \ve X_{\theta_1} {\leftrightarrow} \ve X_{\theta_2} {\leftrightarrow} Y_{\theta_2}$ is Markov and we have 
\begin{equation}
I(\ve X_{\theta_1},\ve X_{\theta_2};Y_{\theta_1},Y_{\theta_2})=I(\ve X_{\theta_1};Y_{\theta_1}) + I(\ve X_{\theta_2};Y_{\theta_2}|Y_{\theta_1}).
\end{equation}
\end{proposition}
\emph{Proof}: The Markovity follows by~\eqref{eq:Ytheta}. We further compute
\begin{align}
I(\ve X_{\theta_1},\ve X_{\theta_2};Y_{\theta_1},Y_{\theta_2}) &\overset{\hphantom{(a)}}=I(\ve X_{\theta_1},\ve X_{\theta_2};Y_{\theta_1}) + I(\ve X_{\theta_1},\ve X_{\theta_2};Y_{\theta_2}|Y_{\theta_1}) \nonumber \\
&= I(\ve X_{\theta_1};Y_{\theta_1}) + I(\ve X_{\theta_2};Y_{\theta_1}|\ve X_{\theta_1})  +I(\ve X_{\theta_2};Y_{\theta_2}|Y_{\theta_1}) +I(\ve X_{\theta_1};Y_{\theta_2}|Y_{\theta_1}, \ve X_{\theta_2}) \nonumber \\
&\overset{(a)}= I(\ve X_{\theta_1};Y_{\theta_1}) + I(\ve X_{\theta_2};Y_{\theta_2}|Y_{\theta_1})
\end{align}
where $(a)$ is consequence of the Markov chain. 
$\hfill \square$

\begin{proposition}
\label{prop:10}
For any $\lambda \geq 0$ we have
\begin{equation}
s_{\lambda}(\ve X_{\theta_1},\ve X_{\theta_2}|T) \geq s_{\lambda}(\ve X_{\theta_1}| T) + s_{\lambda}(\ve X_{\theta_2}| Y_{\theta_1},T)
\end{equation}
with equality if and only we have
\begin{itemize}
\item $I( Y_{\theta_1} ; [\ve X_{\theta_2}]_{\mathcal{S}_m}| [\ve X_{\theta_1}]_{\mathcal{S}_m},T)=0$
\item $I( Y_{\theta_2} ; [\ve X_{\theta_1}]_{\mathcal{S}_m}| Y_{\theta_1},[ \ve X_{\theta_2}]_{\mathcal{S}_m},T)=0$
\end{itemize}
for $m=1,\ldots,M$.
\end{proposition}
\emph{Proof}: We compute   
\begin{align*}
&s_{\lambda}(\ve X_{\theta_1},\ve X_{\theta_2}|T) - s_{\lambda}(\ve X_{\theta_1}| T) - s_{\lambda}(\ve X_{\theta_2}| Y_{\theta_1},T) \\
& = \left(\frac{\lambda}{M-1} +1\right) \left( - I(\ve X_{\theta_1},\ve X_{\theta_2};Y_{\theta_1},Y_{\theta_2}|T) + I(\ve X_{\theta_1};Y_{\theta_1}|T) + I(\ve X_{\theta_2};Y_{\theta_2}|Y_{\theta_1}, T) \right) \\
& + \frac{\lambda}{M-1} \left( \sum_{m=1}^M I([ \ve X_{\theta_1}]_{\mathcal{S}_m},[ \ve X_{\theta_2}]_{\mathcal{S}_m};Y_{\theta_1}, Y_{\theta_2}|T ) - I([ \ve X_{\theta_1}]_{\mathcal{S}_m};Y_{\theta_1}|T ) - I([ \ve X_{\theta_2}]_{\mathcal{S}_m};Y_{\theta_2}|Y_{\theta_1}, T ) \right) \\
&\overset{(a)}{=} \frac{\lambda}{M-1} \sum_{m=1}^M \left( I([\ve X_{\theta_1}]_{\mathcal{S}_m},[\ve X_{\theta_2}]_{\mathcal{S}_m}; Y_{\theta_1}|T) + I([\ve X_{\theta_1}]_{\mathcal{S}_m},[\ve X_{\theta_2}]_{\mathcal{S}_m}; Y_{\theta_2}|Y_{\theta_1},T) - I( [\ve X_{\theta_1}]_{\mathcal{S}_m}; Y_{\theta_1}|T) -I( [\ve X_{\theta_2}]_{\mathcal{S}_m}; Y_{\theta_2}|Y_{\theta_1},T) \right) \\
&= \frac{\lambda}{M-1} \sum_{m=1}^M \left( I([\ve X_{\theta_2}]_{\mathcal{S}_m};Y_{\theta_1}|[\ve X_{\theta_1}]_{\mathcal{S}_m},T) + I( [\ve X_{\theta_1}]_{\mathcal{S}_m}; Y_{\theta_2}|Y_{\theta_1},[\ve X_{\theta_2}]_{\mathcal{S}_m},T)\right) \ge 0
\end{align*}
where $(a)$ follows from Proposition \ref{prop:Mar}. The last step follows from the non-negativity of conditional mutual information terms, and we have equality if and only if all the terms are zero. $\hfill \square$

\begin{proposition} \label{prop:exist}
There is a pair of random variables $(T_*,\ve X_*)$  with $\vert \mathcal{T}_* \vert \leq \frac{J(J+1)}{2}+1$ and $E[\ve X_{*} \ve X_{*}^T] = K$ such that 
\begin{align}
V_{\lambda}(K)=s_{\lambda}(\ve X_{*}|T_{*}).
\end{align}
\end{proposition}

\emph{Proof}: The existence of a maximizer and the cardinality bound on $T_*$ are established in the Appendix \ref{app:E} by using a similar argument as in \cite[Appendix~2A]{Geng--Nair}.
$\hfill \square$


We can now establish the desired result.
\begin{lemma}
\label{lem:2}
Let $p_{*}(t,\ve x)$ attain $V_{\lambda}(K)$ and  let $(T_1,T_2, \ve X_1, \ve X_2) \sim p_{*}(t_1,\ve x_1)p_{*}(t_2,\ve x_2)$. Let $\ve X_t$ denote the conditional distribution $p_*(\ve x|T=t)$ and define
\[
\begin{split}
& \ve X_{\theta_1}|((T_1,T_2)=(t_1,t_2)) \sim \frac{1}{\sqrt{2}}(\ve X_{t_1} + \ve X_{t_2}), \quad   Y_{\theta_1}|((T_1,T_2)=(t_1,t_2)) \sim \frac{1}{\sqrt{2}}( Y_{t_1} +  Y_{t_2}), \\
& \ve X_{\theta_2}|((T_1,T_2)=(t_1,t_2)) \sim \frac{1}{\sqrt{2}}(\ve X_{t_1} - \ve X_{t_2}), \quad   Y_{\theta_2}|((T_1,T_2)=(t_1,t_2)) \sim \frac{1}{\sqrt{2}}( Y_{t_1} - Y_{t_2}). 
\end{split}
\]
Then:
\begin{enumerate}
\item  $(T,  \ve X_{\theta_1})$ also attains $V_{\lambda}(K).$
\item  $(T,  \ve X_{\theta_2})$ also attains $V_{\lambda}(K).$
\item The joint distribution $(T,  \ve X_{\theta_1}, \ve X_{\theta_2})$ must satisfy
\begin{itemize}
\item $I( Y_{\theta_1} ; [\ve X_{\theta_2}]_{\mathcal{S}_m}| [ \ve X_{\theta_1}]_{\mathcal{S}_m},T)=0$
\item $I( Y_{\theta_2} ; [\ve X_{\theta_1}]_{\mathcal{S}_m}| Y_{\theta_1},[ \ve X_{\theta_2}]_{\mathcal{S}_m},T)=0$
\end{itemize}
for $m=1,\ldots,M$.
\end{enumerate}
\end{lemma}

\emph{Proof}: Consider the steps:
\begin{equation}
\label{eqn:pr1}
\begin{split}
2V_{\lambda}(K)&\overset{(a)}{=}s_{\lambda}(\ve X_1|T_1) + s_{\lambda}(\ve X_2|T_2) \\ 
&\overset{(b)}{=}s_{\lambda}(\ve X_1,\ve X_2|T_1,T_2) \\ 
&\overset{(c)}{=}s_{\lambda}(\ve X_{\theta_1},\ve X_{\theta_2}|T_1,T_2)\\
&\overset{(d)}{\geq}  s_{\lambda}(\ve X_{\theta_1}|T_1,T_2) + s_{\lambda}(\ve X_{\theta_2}| Y_{\theta_1},T_1,T_2) \\
&\overset{(e)}{\geq} S_{\lambda}(\ve X_{\theta_1}) + S_{\lambda}(\ve X_{\theta_2}|Y_{\theta_1}) \\
&\overset{(f)}{\geq} S_{\lambda}(\ve X_{\theta_1}) + S_{\lambda}(\ve X_{\theta_2}) \\
&\overset{(g)}{\geq} 2V_{\lambda}(K).
\end{split}
\end{equation}
Here $(a)$ holds for the distribution $p_{*}(t,\ve x)$ that attains $V_{\lambda}(K)$; $(b)$ holds since $(T_1,\ve X_1)$ and $(T_2,\ve X_2)$ are independent by assumption; $(c)$ follows by Proposition \ref{propo:1}; $(d)$ follows by Proposition \ref{prop:10}; $(e)$ follows from 
\begin{align}
s_{\lambda}(\ve X_{\theta_1}|T,Y_{\theta_2})&= \sum_{y_{\theta_2}} p(y_{\theta_2}) s_{\lambda}(\ve X_{\theta_1}| T,Y_{\theta_2}=y_{\theta_2}) \nonumber \\
& \overset{(h)}{\geq}  \sum_{y_{\theta_2}} p(y_{\theta_2}) S_{\lambda}(\ve X_{\theta_1}| Y_{\theta_2}=y_{\theta_2}) \nonumber \\
& \overset{(i)}{=} S_{\lambda}(\ve X_{\theta_1}|Y_{\theta_2}) \label{eqn:defenvelope}
\end{align}
where $(h)$ holds because $S_{\lambda}(\ve X_{\theta_1}|Y_{\theta_2}=y_{\theta_2})$ is the lower convex envelope of $s_{\lambda}(\ve X_{\theta_1}|Y_{\theta_2}=y_{\theta_2})$ and the chain $T \leftrightarrow \ve X_{\theta_1} \leftrightarrow Y_{\theta_1}$ conditioned on $Y_{\theta_2}=y_{\theta_2}$ is Markov (this Markov chain is an immediate implication of (\ref{eq:Ytheta}) where $T=(T_1,T_2)$) and $(i)$ is the definition of $S_{\lambda}(.|.)$;
$(g)$ holds since $S_{\lambda}(\ve X_{\theta_1})$ is convex in $p(\ve x_{\theta_1})$ and by Jensen's inequality $S_{\lambda}(\ve X_{\theta_1}|Y_{\theta_2}) \geq S_{\lambda}(\ve X_{\theta_1})$; $(i)$ follows from definition of $V_{\lambda}(K)$ and by checking the constraint 
\[
\begin{split}
E[\ve X_{\theta_1} \ve X_{\theta_1}^T]&= \sum\limits_{t_1,t_2} p_{*}(t_1)p_{*}(t_2) \frac{1}{2}\left( E[\ve X_{t_1} \ve X_{t_1}^T]+E[\ve X_{t_2} \ve X_{t_2}^T] \right)  \\
&=\sum\limits_{t} p_{*}(t) E[\ve X_{t} \ve X_{t}^T]  = K.
\end{split}
\]
We now see that all inequalities of \eqref{eqn:pr1} are equalities, and step $(d)$ combined with Proposition~\ref{prop:10} proves the claim. $\hfill \square$ \\

\begin{corollary} \label{cor:induc} 
For every $\ell \in \mathbb{N}$ , let $n = 2^{\ell}$ and $(T^n,\ve X_n) \sim \prod_{i=1}^n p_{*}(t_i, \ve x_i)$. Then $(T^n,\tilde{\ve X}_n)$ achieves $V_{\lambda}(K)$ where $\tilde{ \ve X}_n| (T_n = (t_1,t_2,\dots,t_n)) \sim \frac{1}{\sqrt{n}} (\ve X_{t_1} + \ve X_{t_2} +\dots+ \ve X_{t_n})$ .
We choose $\ve X_{t_1} , \ve X_{t_2} , \dots , \ve X_{t_n}$ to be independent random variables.
\end{corollary}
\emph{Proof}: The proof follows by induction using Lemma \ref{lem:2}. $\hfill \square$ \\

Now we state the main Lemma which shows that one of the optimizers is Gaussian. 
\begin{lemma} \label{lemma:stg}
There is a single Gaussian distribution (i.e., the random variable $T$ can be chosen to be a constant) that achieves $V_{\lambda}(K).$
\end{lemma}
\emph{Proof}: See Appendix \ref{app:Fund}. $\hfill \square$ 

Note that our approach does not establish the uniqueness of the maximizing distribution.

\section{Feedback sum-rate capacity for symmetric GMACs}\label{sec:sym}
The proof of Theorem \ref{thm-main} is based on Theorem \ref{thm:gen} with the partition $\mathcal{S}_{1}=\left\{ 1\right\},\mathcal{S}_{2}=\left\{ 2\right\}, \dots, \mathcal{S}_{J}=\left\{ J\right\}$. We tackle the resulting (non-convex) optimization problem with Lagrange duality. 

Consider the covariance matrix
\begin{align}
K=\begin{pmatrix}
Q_1&\rho_{12} \sqrt{Q_1Q_2} & \dots & \rho_{1J} \sqrt{Q_1Q_J}\\
\rho_{21} \sqrt{Q_2Q_1}&Q_2 & \dots &\rho_{2J} \sqrt{Q_2Q_J}\\
\vdots & \vdots & \ddots & \vdots \\
\rho_{J1} \sqrt{Q_JQ_1} & \rho_{J2} \sqrt{Q_JQ_2} & \dots & Q_J
\end{pmatrix}.
\end{align}

\begin{lemma} \label{lemma:multi1}
The original problem is bounded as follows
\begin{align}
-C_{\mathrm{sum}}=\min_{{{\scriptstyle p{(t,\ve x)}:E[X_j^2] \leq P_j } \atop\scriptstyle T \leftrightarrow {\ve X} \leftrightarrow Y }\atop\scriptstyle \text{ subject to } (\ref{eqn:contgen2})} -I(\ve X; Y|T) \geq 
\max_{\lambda} \min_{K \succeq 0: Q_j \leq P_j} & h(\lambda,K)
\end{align}
where
\begin{align}
h(\lambda,K)= \frac{(\lambda-1)}{2} \log \left(1+ \sum_{ j,k=1}^J [K]_{jk} \right) - \frac{\lambda}{2(J-1)} \sum\limits_{j=1}^J \log \left( 1+ \sum_{\ell,k=1}^J [K]_{\ell k} - \frac{\left( \sum_{k=1}^J [K]_{jk} \right)^2}{P_j} \right).
\end{align}
\end{lemma}

\emph{Proof}: See Appendix \ref{app:F}. $\hfill \square$ 

We have shown that the optimal input distributions are Gaussian. At this point the problem is similar to the one in \cite{Wigger--Kim}, and we can use Lemmas 4-5 from \cite{Wigger--Kim} to complete the optimization. However, the converse in \cite{Wigger--Kim} relies on a specific covariance matrix form with only two variables, and this does not necessarily work for asymmetric power constraints. Therefore, we provide a different analysis that applies to asymmetric power constraints.


Consider the covariance matrix
\begin{align}
M=\begin{pmatrix}
P_1&\rho_{12} \sqrt{P_1P_2} & \dots & \rho_{1J} \sqrt{P_1P_J}\\
\rho_{21} \sqrt{P_2P_1}&P_2 & \dots &\rho_{2J} \sqrt{P_2P_J}\\
\vdots & \vdots & \ddots & \vdots \\
\rho_{J1} \sqrt{P_JP_1} & \rho_{J2} \sqrt{P_JP_2} & \dots & P_J
\end{pmatrix}.
\end{align}

\begin{lemma} \label{lemma:multi2}
For every $\lambda \geq 0$, we have
\begin{align}
\max_{\lambda} \min_{K \succeq 0: Q_j \leq P_j} h(\lambda,K) \geq \max_{\lambda} \min\limits_{\scriptstyle \rho_{12}, \dots,  \rho_{N(N-1)}: M \succeq 0}h(\lambda,M).
\end{align}
\end{lemma}
\emph{Proof}: See Appendix \ref{app:G}. $\hfill \square$ \\

\begin{remark}
The argument in Lemma \ref{lemma:multi2} is valid for any dependence balance constraint in \eqref{eqn:ndr}.
\end{remark}

We now set all power constraints to be the same, i.e., $P_1=P_2=\dots=P_J=P$.
\begin{lemma} \label{lemma:multi3}
For every $\lambda \geq 0$, we have
\begin{align} 
\max_{\lambda} \min\limits_{\scriptstyle \rho_{12}, \dots,  \rho_{J(J-1)}: M \succeq 0}h(\lambda,M) \geq \max_{\lambda}\min\limits_{\beta \in [0,J]}\left\{ \frac{(\lambda-1)}{2} \log (1+ JP \beta ) - \frac{J\lambda}{2(J-1)} \log \left( 1+ P \beta \left( J -\beta \right) \right) \right\}.
\end{align}
\end{lemma}
\emph{Proof}: See Appendix \ref{app:H}. $\hfill \square$ \\
\begin{remark}
The paper~\cite{Wigger--Kim} (see also \cite{Kramer}) shows that the optimal covariance matrix is of a simpler form with only two degrees of freedom. Our argument is based on the Cauchy-Schwarz inequality, as shown in the proof of Lemma \ref{lemma:multi3}.
\end{remark}

We define the function
\begin{align}
\ell(\beta,J,P)=\frac{1}{2} \log{(1+ JP\beta)}-\frac{J}{2(J-1)} \log \left( 1+ P \beta \left(J -\beta \right) \right).
\end{align}

\begin{lemma} \label{lemma:multi4}
There exists $\lambda^{*}  \geq 0$ such that
\begin{align} 
&\max_{\lambda} \min\limits_{ \beta \in [0,J]} \left\{ -\frac{1}{2} \log (1+ JP \beta) + \lambda \ell(\beta,J,P) \right\} \\
&=-\frac{1}{2} \log{(1+JP \beta)}
\end{align}
where $\beta \in [1,J]$ is the unique solution to $(1+JP \beta )^{J-1}=\left( 1+ P \beta \left(J -\beta \right) \right)^J$.
\end{lemma}
\emph{Proof}: We have
\begin{align} 
&\max_{\lambda} \min\limits_{ \beta \in [0,J]} \left\{ -\frac{1}{2} \log (1+ JP \beta) + \lambda \ell(\beta,J,P) \right\} \\
&\overset{(a)}{=} \min\limits_{\beta \in [1,J]: \ell(\beta,J,P) \leq 0 }\left\{ -\frac{1}{2} \log (1+ JP\beta) \right\} \\
& \overset{(b)}{=}-\frac{1}{2} \log{(1+JP\beta)}
\end{align}
where $\beta$ is the unique solution satisfying $\beta\in[1,J]$ and $\left(1+PJ\beta \right)^{J-1}=(1+P \beta (J-\beta))^J$; step $(a)$ follows from strong duality as the problem is convex from Lemma \ref{lemma:diff} in Appendix \ref{app:I} and satisfy Slater's condition. Slater's condition holds because there exists a $\beta$ such that $\ell(\beta,J,P) <0$, i.e., $\beta=1$ so the primal problem is strictly feasible. Step $(b)$ follows from the Karush-Kuhn-Tucker (KKT) conditions for a convex problem which satisfy the Slater's condition to find the primal and dual optimal $\beta^{*}$ and $\lambda^{*}$. We start by showing that $\lambda^{*} \neq 0$. Suppose for now that $\lambda^{*}=0$, then from the KKT conditions we have

\begin{align}
\frac{\partial }{\partial \beta} \left\{ -\frac{1}{2} \log (1+ JP \beta) + \lambda \ell(\beta,J,P) \right\} \biggr\rvert_{\lambda=0}= -\frac{JP}{2(1+JP\beta)}=0 \label{eqn:partial}
\end{align} 
which implies that $P=0$. This is impossible, so by contradiction we have $\lambda^{*}\neq 0$. Now by using complementary slackness condition $\lambda^{*} \cdot \ell(\beta^{*},J,P)=0$ we deduce that $\ell(\beta^{*},J,P)=0$, which is equivalent to
\begin{align}
(1+JP \beta^{*})^{J-1}=\left( 1+ P \beta^{*} \left(J - \beta^{*} \right) \right)^J.
\end{align}
This equation has a unique solution for $\beta \in [1,J]$, see~\cite[Lemma~1]{Kramer},~\cite[Appendix~A]{Wigger--Kim}. Finally, the value
\begin{align}
\lambda^{*}= \left( 1- \frac{(J-2 \beta)(1+JP \beta)}{(J-1)(1+P \beta (J-\beta))} \right)^{-1}
\end{align} 
is a valid choice since $\lambda^{*}$ is positive because of \eqref{eqn:partial}, which concludes the proof of the converse. $\hfill \square$
\section{Conclusions} \label{sec:con}
We have derived a new converse bound that combines the Lagrange duality approach of \cite{Kramer--Michael2006} with a novel variation of the factorization of a convex envelope \cite{Geng--Nair}. The new converse bound meets the achievable sum-rate of the Fourier MEC scheme, thus establishing the sum-rate capacity for the $J$-user symmetric GMAC with feedback.

It remains to see whether Fourier MEC can achieve all rate points in the capacity region of the symmetric GMAC with feedback. For asymmetric transmit power constraints, however, it is known that Fourier-MEC can be improved by using modulation frequencies other than the uniformly-spaced frequencies $(j-1)/J$ for $j\in\mathcal{J}$. A few more variations of MEC strategies are described in~\cite[Sec.~VIII]{Kramer}. 

\begin{example}
Fourier MEC does not meet the dependence-balance bound under asymmetric power constraints. For example, consider three users with the power constraints $P_1 \leq 1, P_2 \leq 4$ and $P_3 \leq 9$, for which Fourier-MEC achieves the sum-rate $R_\text{sum}=1.6215$ bits/use, see~\cite[Sec.~III]{Kramer}. However, the dependence balance bound permits a larger sum-rate, since the choice $(\rho_{12},\rho_{13},\rho_{23})=(0.5,0.44,0.58)$ satisfies the dependence balance constraints and permits $R_\text{sum}=1.6427$.
\end{example}

\appendices

\section{Existence of minimizing distribution} \label{app:E}
%




\begin{proposition}[{\cite[Lemma~1]{Boos}}] \label{prop:strgcon}
Suppose that $ Y_n$ and $ Y$ have continuous densities $f_{n}( y)$, $f( y)$ with respect to the Lebesgue measure on $\mathbb{R}$. If $Y_n \overset{w}{\Rightarrow} Y$ and
\begin{align}
\sup_{n}\vert f_{n}({y}) \vert \leq M({ y})<\infty,~\forall{y}\in \mathbb{R}
\end{align}
and $f_{n}~{\rm is~equicontinuous,~i.e.,}~\forall~{y},\epsilon>0,~\exists \delta ({y},\epsilon), n({y},\epsilon)$ such that $\Vert y- y_1 \Vert < \delta(y, \epsilon)$ implies that $\vert f_n( y)-f_n(y_1) \vert < \epsilon$ $\forall n \geq n( y, \epsilon)$, then for any compact subset $C$ of $\mathbb{R}$ we have
\begin{align}
\sup_{{y}\in C}\vert f_{n}({ y})-f({y})\vert\rightarrow 0~{\rm as}~n\rightarrow\infty.
\end{align}

If $\left\{f_n\right\}$ is uniformly equicontinuous, i.e., $\delta( y,\epsilon)$, $n( y,\epsilon)$ do not depend on $ y$, and $f( y_n) \rightarrow 0$ whenever $\Vert  y_n\Vert \rightarrow \infty$ then
\begin{align}
\sup_{{ y}\in \mathbb{R}}\vert f_{n}({ y})-f({ y})\vert=\Vert f_{n}({ y})-f({ y})\Vert_{\infty}\rightarrow 0~{\rm as}~n\rightarrow\infty.
\end{align}
\end{proposition}

\begin{proposition}[{\cite[Proposition~16]{Geng--Nair}}]
Let $\left\{ \ve X_n\right\}$ be any sequence of random variables satisfying $ Y_n=G\ve X_n+\ve Z$ where $\ve Z \sim \mathcal{N}(0,I)$ is independent of $\left\{ \ve X_n\right\}$ and $f_n(y)$ represent the density of $Y_n$. Then the collection of functions $\left\{f_n( y)\right\}$ is uniformly bounded and uniformly equicontinuous.
\end{proposition}

\begin{definition}
A collection of random variables $\ve X_n$ on $\mathbb{R}^N$ is said to be tight if for every $\epsilon>0$ there is a compact set $C_{\epsilon} \subset \mathbb{R}^N$ such that $P(\ve X_n \not\in C_{\epsilon}) \leq \epsilon$, $\forall n$.
\end{definition}

%
\begin{proposition}[{\cite[Proposition~17]{Geng--Nair}}] \label{prop:tight}
Consider a sequence of random variables $\left\{\ve X_n\right\}$ that satisfies the covariance constraint $E[\ve X_n \ve X^T_n] = K$, $\forall n$. Then the sequence is tight.
\end{proposition}

\begin{theorem}[Prokhorov] \label{thm:prok}
If $\left\{\ve X_n\right\}$ is a tight sequence of random variables in $\mathbb{R}^N$ then there exists a subsequence $\left\{\ve X_{n_{i}}\right\}$ and a limiting probability distribution $\ve X_{*}$ such that $\ve X_{n_{i}} \overset{w}{\Rightarrow} \ve X_{*}$.
\end{theorem}

%

\begin{proposition}[{\cite[Proposition~18]{Geng--Nair}}] \label{prop:entrp}
Let $\ve X_{n} \overset{w}{\Rightarrow} \ve X_{*}$ and let $Z \sim \mathcal{N}(0,1)$ be pairwise independent of $\left\{ \ve X_n\right\}$, $\ve X_{*}$. Let $ Y_n=G \ve X_n+ Z$, $Y_{*}=G\ve X_{*}+Z$. Further let $E[\ve X_n \ve X^T_n] = K$, $E[\ve X_{*} \ve X^T_{*}] = K$. Let $f_n( y)$ denote the density of $Y_n$ and $f_{*}(y)$ denote the density of $ Y_{*}$. Then we have
\begin{enumerate}
\item $Y_{n} \overset{w}{\Rightarrow} Y_{*}$, 
\item $f_n( y) \rightarrow f_{*}( y)$ for all $y$, 
\item $h( Y_n) \rightarrow h(Y_{*})$.
\end{enumerate}
\end{proposition}

\begin{proposition}[Lower Semi-continuity] \label{prop:entroprim}
Let $\ve X_{n} \overset{w}{\Rightarrow} \ve X_{*}$ and $ Y_n=G \ve X_n+ Z$, $Y_{*}=G\ve X_{*}+Z$, where $Z \sim \mathcal{N}(0,1)$ is pairwise independent of $\left\{ \ve X_n\right\}$, $\ve X_{*}$. Let $s_{\lambda}(\ve X_n)=(\lambda-1)h(Y_n)+\left( \frac{\lambda}{J-1} +1 \right)h(Z) -\frac{\lambda}{J-1}\sum_{j=1}^J h(Y_n|X_{jn})$ and $s_{\lambda}(\ve X_{*})$ similarly. Then
\begin{enumerate}
\item $(Y_{n},X_{1n}) \overset{w}{\Rightarrow} (Y_{*},X_{1*})$, 
\item $\lim\inf_{n \leftrightarrow \infty} s_{\lambda}(\ve X_n) \geq s_{\lambda}(\ve X_{*})$. 
\end{enumerate}
\end{proposition}

\emph{Proof}: The first part follows from pointwise convergence of characteristic functions (which is equivalent to weak convergence by Levy's continuity theorem) since $\Phi_{(\ve X_n, Z )}(\ve u,v) =E[e^{i \ve u \ve X_n+ i v Z}] = E[e^{i \ve u \ve X_n}] E[e^{i v Z}] =\Phi_{\ve X_n}(\ve u)\Phi_{Z}(v)$ then by letting $n \rightarrow \infty$ we have $\Phi_{\ve X_*}(\ve u)\Phi_{Z}(v)= E[e^{i \ve u \ve X_{*}}] E[e^{i v Z}] = E[e^{i \ve u \ve X_n+ i v Z}] = \Phi_{(\ve X_{*}, Z)}(\ve u, v)$. To relate $(Y_n, X_{1n})$ with $(\ve X_n, Z)$ we use the linear transformation $(Y_n, X_{1n})^T=A(\ve X_n, Z)^T$ for a deterministic matrix $A$.
By using the previous steps and the linear dependence we obtain $\lim_{n \rightarrow \infty} \Phi_{(Y_n, X_{1n})}(\ve t)= \lim_{n \rightarrow \infty}\Phi_{(\ve X_n, Z)}(A\ve t) =\Phi_{(X_{*}, Z)}(A\ve t)= \Phi_{(Y_{*}, X_{1*})}(\ve t)$. 

For the second part fix $\delta>0$ and define $N_{\delta} \sim \mathcal{N}(0,\delta)$, pairwise independent of $\left\{ \ve X_n \right\}$, $\ve X_{*}$. From the Markov chain $X_{1n}+N_{\delta} \leftrightarrow X_{1n} \leftrightarrow Y_n$ and the data processing inequality we have $h(Y_n|X_{1n}) \leq h(Y_n|X_{1n}+N_{\delta})$. By the third claim of proposition \ref{prop:entrp}, we obtain 
\begin{align}
(\lambda-1)h(Y_n)+\left( \frac{\lambda}{J-1} +1 \right)h(Z) -\frac{\lambda}{J-1}\sum_{j=1}^J h(Y_n|X_{jn}+J_{\delta}) \rightarrow
(\lambda-1)h(Y_*)+\left( \frac{\lambda}{J-1} +1 \right)h(Z) -\frac{\lambda}{J-1}\sum_{j=1}^J h(Y_*|X_{j*}+N_{\delta})
\end{align}
as $n \leftrightarrow \infty$. Thus, we have
\begin{align} \label{eqn:semiconvetgence}
\lim\inf_{n \leftrightarrow \infty} s_{\lambda}(\ve X_n) \geq (\lambda-1)h(Y_*)+\left( \frac{\lambda}{J-1} +1 \right)h(Z) -\frac{\lambda}{J-1}\sum_{j=1}^J h(Y_*|X_{j*}+N_{\delta}).
\end{align}
Since the RHS of (\ref{eqn:semiconvetgence}) is continuous in $\delta$, we take $\delta \downarrow 0$ and prove the second claim. $\hfill \square$

\begin{theorem}[{\cite[Theorem~1]{Hero}}] \label{thm:concon}
Let $\left\{ Y_i \in \mathcal{C} \right\}$ be a sequence of continuous random variables with pdf's $\left\{f_i\right\}$ and $ Y_{*}$ be a continuous random variable with pdf $f_{*}$ such that $f_{i} \rightarrow f_{*}$ pointwise. Let $\Vert y\Vert=\sqrt{ y^\dag y}$ denote the Euclidean norm of $ y \in \mathcal{C}$. If the conditions

\begin{align}
\max\{\sup_{y}f_{i}({y}),\sup_{y}f_{\ast}({ y})\}\leq & F, \\
\max\{\int\Vert{ y}\Vert^{\kappa}f_{i}({y}) d{y},\int\Vert{y}\Vert^{\kappa}f_{\ast}({y}) d{y}\} \leq & L
\end{align}
hold for some $\kappa >1$ and for all i then $h(Y_i) \rightarrow h(Y_{*})$.
\end{theorem}

\begin{remark}
This theorem is relatively straightforward. We have $\lim\inf_i h( Y_i) \geq h(Y_{*})$ due to the upper bound on the densities and $\lim\sup_i h(Y_i) \leq h(Y_{*})$ due to the moment constraints.
\end{remark}

\emph{Proof of Proposition \ref{prop:exist}}: Define

\begin{align}
{v}_{\lambda}({\hat{K}})=\inf_{ p(\ve x):E [{\ve X}{\ve X}^{T}]={\hat{K}}}{s}_{\lambda}({\ve X}).
\end{align}
Let $\ve X_n$ be a sequence of random variables such that $E[\ve X_n \ve X^T_n]=\hat{K}$  and $s_{\lambda}(\ve X_n) \downarrow v_{\lambda}(\hat{K})$. By the covariance constraint (Proposition \ref{prop:tight}) we know that the sequence of random variables $\ve X_n$ forms a tight sequence and by Theorem \ref{thm:prok} there exists $X^{*}_{\hat{K}}$ and a convergent subsequence such that $\ve X_{n_{i}} \overset{w}{\Rightarrow} \ve X^{*}_{\hat{K}}$. From Proposition \ref{prop:entrp} and \ref{prop:entroprim} we have $s_{\lambda}(\ve X^{*}_{\hat{K}})=v_{\lambda}(\hat{K} )$. For $\lambda \geq 0$ we have the following trivial bound

\begin{align}
{ v}_{\lambda}({\hat{K}})={ s}_{\lambda}({\ve X}_{\hat{K}}^{\ast})\geq -\left(\frac{\lambda}{J-1}+1\right) I({\ve X}_{\hat{K}}^{\ast};{Y}) \geq \left( {-{\lambda+J-1}\over{2(J-1)}} \right)\log( 1+G{\hat{K}}G^{T})=C_{\lambda}.
\end{align}

Recall that $V_{\lambda}(K)$ is defined using a convex combination:
\begin{align}
{ V}_{\lambda}(K)=\inf_{{\scriptstyle (V,{\ve X}):E[{\ve X}{\ve X}^{T}]=K}\atop\scriptstyle T\leftrightarrow{\ve X}\leftrightarrow Y}{ s}_{\lambda}({\ve X}\vert V).
\end{align}
Hence to obtain the best convex combination subject to the covariance constraint it suffices to consider the family of maximizers $\ve X^{*}_{\hat{K}}$ for $\hat{K} \succeq 0$. Thus, we have
\begin{align}
{ V}_{\lambda}(K)=\inf_{{\scriptstyle\alpha_{i},{\hat{K}}_{i}:\alpha_{i}\geq0,\sum\limits_{i}\alpha_{i}=1}\atop\scriptstyle\sum\limits_{i}\alpha_{i}{\hat{K}}_{i}= K}\sum\nolimits_{i}\alpha_{i}{ v}_{\lambda}({\hat{K_{i}}}).
\end{align}
It takes $\frac{J(J+1)}{2}$ constraints to preserve the covariance matrix and one constraint to preserve $\sum_{i} \alpha_i v_{\lambda} (\hat{K}_i)$. Hence, by using the Bunt-Caratheodory theorem, we can consider convex combinations of at most $m:=\frac{J(J+1)}{2}+1$ points, i.e.,
\begin{align}
{V}_{\lambda}(K)=\inf_{{\scriptstyle\alpha_{i},{\hat{K}}_{i}:\alpha_{i}\geq0,\sum\nolimits_{i=1}^{m}\alpha_{i}=1}\atop\scriptstyle\sum_{i=1}^{m}\alpha_{i}{\hat{K}}_{i}= K}\sum\limits_{i=1}^{m}\alpha_{i}{ v}_{\lambda}({\hat{K_{i}}}).
\end{align}

Consider any sequence of convex combinations $\left(\left\{\alpha_i^n \right\},\left\{K_i^n \right\} \right)$ that approaches the supremum as $n \rightarrow \infty$. Using compactness of the $m-$dimensional simplex, we can assume w.l.o.g. that $ \alpha_i^n \overset{n \rightarrow \infty}{\rightarrow} \alpha_i^{*}$, $i=1,\dots,m$. If any $\alpha_i^{*}=0$, since $\alpha_i^{n}K_i^{n} = K$ and $v_{\lambda}(K_i^{n}) \geq C_{\lambda}$ it is easy to see that $\alpha_i^{n} v_{\lambda}(K_i^{n}) \overset{n \rightarrow \infty}{\rightarrow} 0$. Thus we can assume that $\min_{i=1,\dots,m}\alpha_i^{*}=\alpha^{*}>0$. This implies that $K_i^n \preceq \frac{2}{\alpha^{*}K}$ for large enough $n$ uniformly in $i$. Hence we can find a convergent subsequence for each $i$, $1 \leq i \leq m$, so that $K_i^{n_k} \overset{k \rightarrow \infty}{\rightarrow} K_i^{*}$. We thus have
\begin{align}
{ V}_{\lambda}(K)=\sum_{i=1}^{m}\alpha_{i}^{\ast}{ v}_{\lambda}({\hat{K_{i}^{\ast}}}).
\end{align}
In other words, we can find a pair of random variables $(T_{*},\ve X_{*})$ with $\vert \mathcal{T} \vert \leq \frac{J(J+1)}{2}+1$ such that $V_{\lambda}(K)=s_{\lambda}(\ve X_{*}|T_{*})$. $\hfill \square$


\section{Proof of Lemma \ref{lemma:stg}} \label{app:Fund}
We define the set of typical sequences as $\mathcal{T}^{(n)}(T):= \left\{ t^n : \big\vert \lvert \left\{ i:t_i =t \right\} \rvert - n p_{*}(t) \big\vert \leq n \omega_n p_{*}(t), \forall t \in [1 : m] \right\},$ where $\omega_n$ is any sequence such that $\omega_n \rightarrow 0$ as $n \rightarrow \infty$ and $\omega_n \sqrt{n} \rightarrow \infty$ as $n \rightarrow \infty$. For instance $\omega_n = \frac{\log{n}}{\sqrt{n}}$ . By using Chebyshev's inequality, we have
\begin{align*}
P(\big\vert \lvert \left\{ i:t_i =t \right\} \rvert - n p_{*}(t) \big\vert \leq n \omega_n p_{*}(t)) \leq \frac{1-p_{*}(t)}{p_{*}(t) \omega_n^2 n}.
\end{align*}
Hence $P(t^n \notin \mathcal{T}^{(n)}(T)) \rightarrow 0$ as $n \rightarrow \infty$. Consider a sequence of induced distributions $\hat{\ve X}_n \sim \tilde{\ve X}_n|t^n$, where $\tilde{\ve X}_n|t^n$ and $\tilde{\ve X}_n|(T^n=t^n)$ denote the same thing.
\begin{proposition} \label{prop:geg}
$\hat{\ve X}_n \overset{w}{\Rightarrow} \mathcal{N}(0,\sum_{t=1}^m p_{*}(t)K_t)$.
\end{proposition}
\emph{Proof}: For given $t^n$, let $A_n(t)=| \left\{ i: t_i=t \right\}|$. We know that $A_n(t) \in np_{*}(t)(1\pm \omega_n)$, $\forall t$. Consider a $\ve c$ with entries in real number and $\Vert \ve c \Vert =1$. Let $\hat{ \ve X}_{n,i}^{\ve c} \sim \frac{1}{\sqrt{n}} \ve c^T \cdot \ve X_{t_i}$ and $\hat{ \ve X}_{n,i}^{\ve c}$ be independent random variable over $i$. Observe that $\sum_{i=1}^n \hat{ \ve X}_{n,i}^{\ve c} \sim \ve c^T \hat{\ve X}_n$. Note that 

\begin{align*}
\sum_{i=1}^n E[(\hat{ \ve X}_{n,i}^{\ve c})^2] &= \frac{1}{n} \sum_{t} A_n(t) \ve c^T K_t \ve c \\
& \rightarrow \ve c^T \left(\sum_{t} p_{*}(t)K_t \right) \ve c.
\end{align*}
\begin{align*}
\sum_{i=1}^n E[(\hat{ \ve X}_{n,i}^{\ve c})^2; |\hat{ \ve X}_{n,i}^{\ve c}| > \epsilon_1] &= \frac{1}{n} \sum_{t} A_n(t) E[\ve c^T \ve X_t \ve X_t^T \ve c; \ve c^T \ve X_t \ve X_t^T \ve c >n \epsilon_1^2] \\
& \leq  \sum_{t} p_{*} (t) (1+\omega_n) E[\ve c^T \ve X_t \ve X_t^T \ve c; \ve c^T \ve X_t \ve X_t^T \ve c >n \epsilon_1^2] \rightarrow 0.
\end{align*}

For the last step, we used that the $K_t$'s are bounded, and hence $\ve c^T \ve X_t$ has a bounded second moment. The Lindeberg-Feller Central Limit Theorem gives $\sum_{i=1}^n \hat{ \ve X}_{n,i}^{\ve c} \overset{w}{\Rightarrow} \mathcal{N}(0, \ve c^T \sum_t p_{*}(t) K_t \ve c)$. Hence $\hat{ \ve X}_{n} \overset{w}{\Rightarrow} \mathcal{N}(0, \sum_t p_{*}(t) K_t)$ from Kramer-Wold. $\hfill \square$

\begin{proposition} \label{prop:conti}
Given any $\delta>0$, there exists $N_0$ such that $\forall n > N_0$ we have for all $t^n \in \mathcal{T}^{(n)}(T)$
\begin{align*}
s_{\lambda}(\tilde{\ve X}_n|t^n) - s_{\lambda}(\ve X^{*}) \leq \delta,
\end{align*}
where $\ve X^{*} \sim \mathcal{N}(0,\sum_{t} p_{*}(t)K_t)$.
\end{proposition}
\emph{Proof}: Assume the claim is not true. Then we have a subsequence $t^{n_k} \in \mathcal{T}^{n_k}(T)$ and distributions $\tilde{\ve X}_{n_k}| t^{n_k}$ such that
\begin{align*}
s_{\lambda}(\tilde{\ve X}_{n_k}|t^{n_k}) > s_{\lambda}(\ve X^{*})+ \delta, \forall k.
\end{align*}
However from Proposition \ref{prop:geg} we know that $\tilde{\ve X}_{n_k}| t^{n_k} \overset{w}{\Rightarrow} \ve X^{*}$ and from Proposition \ref{prop:entrp} we have $s_{\lambda}(\tilde{\ve X}_{n_k}|t^{n_k}) \rightarrow s_{\lambda}(\ve X^{*})$, a contradiction. $\hfill \square$

\emph{Proof of Lemma \ref{lemma:stg}}: We know from Corollary \ref{cor:induc} that for every $\ell \in \mathbb{N}$ and $n=2^{\ell}$, the pair $(T^n,\tilde{\ve X}_n)$ achieves $V_{\lambda}(K)$. Hence we have
\begin{align*}
V_{\lambda}(K) &=\sum_{t^n} p_{*} (t^n) s_{\lambda}(\tilde{\ve X}_n|t^n) \\
&=\sum_{t^n \in \mathcal{T}^{(n)}(T)} p_{*} (t^n) s_{\lambda}(\tilde{\ve X}_n|t^n) + \sum_{t^n \notin \mathcal{T}^{(n)}(T)} p_{*} (t^n) s_{\lambda}(\tilde{\ve X}_n|t^n).
\end{align*} 
For a given $t^n$, let $\hat{\ve X} \sim \tilde{\ve X}_n|t^n$ so that $E[\hat{\ve X}\hat{\ve X}^T] \preceq \sum_{t=1}^m K_t$. Thus $s_{\lambda}(\hat{\ve X}) \leq C_{\lambda}$ for some fixed constant that is independent of $t^n$. Using Proposition \ref{prop:conti} we can upper bound $V_{\lambda}(K)$ for large $n$ by

\begin{align*}
V_{\lambda}(K) &=\sum_{t^n \in \mathcal{T}^{(n)}(T)} p_{*} (t^n) s_{\lambda}(\tilde{\ve X}_n|t^n) + \sum_{t^n \notin \mathcal{T}^{(n)}(T)} p_{*} (t^n) s_{\lambda}(\tilde{\ve X}_n|t^n) \\
& \leq \sum_{t^n \in \mathcal{T}^{(n)}(T)} p_{*} (t^n) (s_{\lambda}(\ve X^{*})+\delta) + C_{\lambda}\sum_{t^n \notin \mathcal{T}^{(n)}(T)} p_{*} (t^n) \\
& =P(t^n \in \mathcal{T}^{(n)}(T))(s_{\lambda}(\ve X^{*})+\delta) + C_{\lambda}P(t^n \notin \mathcal{T}^{(n)}(T)).
\end{align*} 

Here $\ve X^{*} \sim \mathcal{N}(0, \sum_{t} p_{*}(t)K_t)$. Since $P(t^n \in \mathcal{T}^{(n)}(T)) \rightarrow 1$ as $n \rightarrow \infty$ we have $V_{\lambda}(K) \leq s_{\lambda}(\ve X^{*}) + \delta$. However, $\delta>0$ is arbitrary, and hence $V_{\lambda}(K) \leq s_{\lambda}(\ve X^{*})$. The other direction $V_{\lambda}(K) \geq s_{\lambda}(\ve X^{*})$ follows from the definition of $V_{\lambda}(K)$ and $\sum_{t} p_{*}(t) K_t \preceq K$. $\hfill \square$

\section{Proof of Lemma \ref{lemma:multi1}} \label{app:F}
By the standard Lagrangian duality we have
\begin{equation}
\begin{aligned}
&-C_{\mathrm{sum}}= \min_{\substack{ p(t,\ve x):~E[X_j^2] \le P_j  \\ T {\leftrightarrow} {\ve X} {\leftrightarrow} Y \\ \text{ subject to } \eqref{eqn:bes}}} -I(\ve X; Y|T) \overset{(a)}{\geq} \max_{\lambda} \min_{\substack{p(\ve x) \in G_{\mathcal{G}}}} s_{\lambda}(\ve X) \\[5pt]
&= \max_{\lambda} \min_{K \succeq 0: Q_j \leq P_j} \left\{ \frac{(\lambda-1)}{2} \log \left(1+ \sum_{ j,k=1}^J [K]_{jk} \right) - \frac{\lambda}{2(J-1)} \sum\limits_{j=1}^J \log \left( 1+ \sum_{\ell,k=1}^J [K]_{\ell k} - \frac{\left( \sum_{k=1}^J [K]_{jk} \right)^2}{P_j} \right)  \right\}
\end{aligned} 
\end{equation}

where $(a)$ follows from Theorem \ref{thm:gen} and last step by inserting an optimal Gaussian \emph{input} distribution. $\hfill \square$ 

\section{Proof of Lemma \ref{lemma:multi2}} \label{app:G}  
For the Gaussian MAC with $g_j=1$ for all $j$ we have
\begin{align}
K_{\ve X Y}=\begin{pmatrix}
K_{\ve X}& \Cov{(\ve X,Y)} \\
\Cov{( \ve X,Y)}^T  &K_Y
\end{pmatrix}=\begin{pmatrix}
K_{\ve X}& K_{\ve X} \ve 1 \\
(K_{\ve X} \ve 1)^T  & 1+ \ve 1^T K_{\ve X} \ve 1
\end{pmatrix} \label{eq:ind}
\end{align}
where $K_Y=1+ \ve 1^T K_{\ve X} \ve 1$, $\Cov{(\ve X,Y)}=K_{\ve X} \ve 1$ and $\ve 1$ is column vector of all ones.
The function $h(\cdot)$ can be rewritten as follows:
\begin{align}
2h(\lambda, K_{\ve X})=- \log{\det{K_{Y}}} - \frac{\lambda}{J-1} \log \frac{\prod\limits_{j=1}^J \det{K_{X_j Y}}}{ \left( \det{K_Y} \right)^{J-1} \prod\limits_{j=1}^J \det{K_{X_j}}}.
\end{align}
We define $K^{\prime}_{\ve X}$ to be the same as $K_{\ve X}$ except that the (1,1) entry of $K^{\prime}_{\ve X}$ is $P_1$ rather than $Q_1$. Then from \eqref{eq:ind} we have
\begin{align}
&K^{\prime}_{X_1 Y}= K_{X_1 Y} \circ \begin{pmatrix}
D & F \\
F & E\\
\end{pmatrix}, \\
&K^{\prime}_{X_j Y}= K_{X_j Y} \circ \begin{pmatrix}
1 & 1 \\
1 & E\\
\end{pmatrix}, \quad j  \neq 1
\end{align}
where `$\circ$' denotes Hadamard multiplication, and where
\begin{align}
D= \frac{P_1}{Q_1}, \quad E= \frac{K_Y+P_1 -Q_1}{K_Y}, \quad F=\frac{\Cov{(X_1,Y)}+P_1-Q_1}{\Cov{(X_1,Y)}}.
\end{align}  
Observe that $D>1$ and $E>1$. Now by using Oppenheim's inequality ($\det K^{\prime}_{X_1 Y} \geq D E \det K_{X_1 Y}$) \cite[p. 480]{Horn--Jonson} we have
\begin{align}
2h(\lambda, K^{\prime}_{\ve X})&= - \log{E K_Y} - \frac{\lambda}{J-1} \log \frac{ \left( \prod\limits_{j=2}^J \det K^{\prime}_{X_j Y} \right) \det K^{\prime}_{X_1 Y}}{E^{J-1} DK_Y^{J-1} \prod\limits_{j=1}^J K_{X_j}} \\
& \leq - \log{E K_Y} - \frac{\lambda}{N-1} \log E\frac{\prod\limits_{j=1}^J \det K_{X_j Y}}{K_Y^{J-1} \prod\limits_{j=1}^J K_{X_j}} \\
&= -(1+\frac{\lambda}{J-1})\log{E} + 2h(\lambda, K_{\ve X})  \leq  2h(\lambda, K_{\ve X}).
\end{align}

The minimum is reached for $K^{\prime}_{\ve X}$ which is a covariance matrix from the argument below
\begin{align}
K^{\prime}_{\ve X}=\begin{pmatrix}
P_1 &\rho_{12} \sqrt{Q_1Q_2} & \dots & \rho_{1J} \sqrt{Q_1Q_J}\\
\rho_{21} \sqrt{Q_2Q_1}&Q_2 & \dots &\rho_{2J} \sqrt{Q_2Q_J}\\
\vdots & \vdots & \ddots & \vdots \\
\rho_{J1} \sqrt{Q_JQ_1} & \rho_{J2} \sqrt{Q_JQ_2} & \dots & Q_J
\end{pmatrix}=\begin{pmatrix}
P_1 &\rho_{12} \sqrt{\frac{Q_1}{P_1}} \sqrt{P_1Q_2} & \dots & \rho_{1J} \sqrt{\frac{Q_1}{P_1}} \sqrt{P_1Q_J}\\
\rho_{21} \sqrt{\frac{Q_1}{P_1}} \sqrt{Q_2P_1}&Q_2 & \dots &\rho_{2J} \sqrt{Q_2Q_J}\\
\vdots & \vdots & \ddots & \vdots \\
\rho_{J1} \sqrt{\frac{Q_1}{P_1}} \sqrt{Q_JP_1} & \rho_{J2} \sqrt{Q_JQ_2} & \dots & Q_J
\end{pmatrix}.
\end{align}
%
With the same approach we attain the minimum for $Q_2=P_2$, $Q_3=P_3$, $\dots$, $Q_J=P_J$. Thus, we can get the desired lower bound on the original problem.


\section{Proof of Lemma \ref{lemma:multi3}} \label{app:H}

Consider the arithmetic mean
\begin{align}
\rho = \frac{1}{J(J-1)} \left( \sum\limits_{{\scriptstyle j,k=1} \atop\scriptstyle j \neq k}^J \rho_{jk} \right).
\end{align}

For the inequality in Lemma \ref{lemma:multi3} to hold we need to show the following inequality
\begin{align}
\prod_{j=1}^J \left( 1+\sum_{\ell,k=1}^J M_{\ell k} - \frac{\left(\sum_{k=1}^J M_{jk} \right)^2}{P}  \right) \leq \left( 1+ P \beta \left(J -\beta \right) \right)^J.
\end{align}

We prove it as follows:
\begin{align}
\prod_{j=1}^J \left( 1+\sum_{\ell,k=1}^J M_{\ell k} - \frac{\left(\sum_{k=1}^J M_{jk} \right)^2}{P}  \right)
&\overset{(a)}{\leq} \left( 1+\sum_{\ell,k=1}^J M_{\ell k} - \sum\limits_{j=1}^J \frac{\left( \sum\limits_{k=1}^J M_{jk} \right)^2}{JP} \right)^J \\
&\overset{(b)}{\leq} \left( 1+\sum_{\ell,k=1}^J M_{\ell k} - \left( \sum\limits_{j,k=1}^J \frac{M_{jk}}{J\sqrt{P}} \right)^2 \right)^J \\
&= \left( 1+ P\beta \left( J-\beta \right) \right)^J \label{eqn:compactform}
\end{align}
where $(a)$ follows from arithmetic-geometric mean (AM-GM) 
which is valid for non-negative real numbers, and $(b)$ follows from Cauchy-Schwarz inequality 
\begin{align}
(1^2+1^2+ \dots+1^2) \left( \sum\limits_{j=1}^J \left( \sum\limits_{k=1}^J \frac{M_{jk}}{\sqrt{P}} \right)^2 \right) \geq \left( \sum\limits_{j,k=1}^J \frac{M_{jk}}{\sqrt{P}} \right)^2.
\end{align}
For equality in both $(a)$ and $(b)$ a sufficient and necessary condition is $\rho_{12}=\rho_{13}= \dots=\rho_{(J-1)J}=\rho$. Define $\beta=1+(J-1)\rho$, therefore to obtain the expression in (\ref{eqn:compactform}) we will use the identity $\sum_{\ell, k=1}^J M_{\ell k}= PJ \beta$. Since $-1/(J-1) \leq \rho \leq 1$, we have $0 \leq \beta \leq J$.
%
%
%

\section{Convexity} \label{app:I}
\begin{lemma} \label{lemma:diff}
The problem $\min\limits_{\beta \in [0,J]: \ell(\beta,J,P) \leq 0 }\left\{ -\frac{1}{2} \log (1+ JP \beta) \right\}$ is convex, where 
\begin{align}
\ell(\beta,J,P)=\frac{1}{2} \log{(1+JP \beta)}-\frac{J}{2(J-1)} \log{\left( 1+ P \beta \left( J - \beta \right) \right)}.
\end{align}
\end{lemma}

\emph{Proof}: 
For a fixed $P$ and $J$, the term $-\log (1+ JP \beta)$ is convex on $\beta$. We now show that, for a fixed $P$ and $J$, $\ell(\beta,J,P)$ is convex in $\beta$. We split the derivations into two steps. In the first step, we show that $\frac{\partial^2\ell(\beta,J,P)}{\partial \beta^2} \geq 0$ for $\beta \geq 1$. In the second step, we show that $\beta \in  \left[ 1, J \right]$ are the only possible maximizers for the original problem.
\begin{enumerate}
\item We compute $\frac{\partial^2\ell(\beta,J,P)}{\partial \beta^2}$ as
\begin{align}
\begin{pmatrix}
1 & \gamma & \gamma^2 & \gamma^3 & \gamma^4
\end{pmatrix}
\begin{pmatrix}
M^3P^4 + 8M^2P^4 + 2M^2P^3 + 22MP^4 + 14MP^3 \\
+ 3MP^2 + 21P^4 + 24P^3 + 11P^2 + 2P \\ \\
2M^3P^4 + 16M^2P^4 + 2M^2P^3 + 46MP^4 + 20MP^3 \\
+ 2MP^2 + 48P^4 + 42P^3 + 10P^2 \\ \\
M^3P^4 + 9M^2P^4 + 33MP^4 + 10MP^3 + 45P^4 \\
+ 30P^3 + 2P^2 \\ \\
2M^2P^4 + 16MP^4 + 4MP^3 + 30P^4 + 12P^3 \\ \\
M^2P^4 + 7MP^4 + 12P^4 
\end{pmatrix} >0 \nonumber
\end{align} 
where $M=J-3 \geq 0$ and $\gamma=\beta-1 \geq 0$ for the case of three or more users ($J \ge 3$). For $P>0$ and $\gamma \geq0$ the above inequality holds from the addition of only positive terms, thus the problem is convex. To show that the problem is convex for all admissible values of $\beta$, we need to rule out some values and that is done in the next part. 
\item Assume that $\beta_0 \in \left[0, 1 \right[$ is the minimizer of the problem defined above, so that the following holds
\begin{align}
\min\limits_{\beta \in [0,J]: \ell(\beta,J,P) \leq 0}\left\{ -\frac{1}{2} \log (1+ JP \beta) \right\}= -\frac{1}{2} \log (1+ JP \beta_0).
\end{align}
For $\beta_1=1 $, the constraint $\ell(1,J,P)<0$ is satisfied, which is equivalent to showing that 
\begin{align} \label{eq:depbalineq}
(1+JP)^{J-1} < (1+(J-1)P)^J
\end{align}
for $P>0$. But we have
\begin{align}
\left( \frac{1+(J-1)P}{1+JP} \right)^{J-1} &= \left( 1- \frac{P}{1+JP} \right)^{J-1} \\
& \geq \frac{2}{1+JP}  \label{eq:Bernoulli} \\
& > \frac{1}{1+(J-1)P}
\end{align}
where \eqref{eq:Bernoulli} follows by Bernoulli's inequality ($(1+x)^r \geq 1+rx$ for $r\geq1,x \geq -1$). The new minimizer is $\beta_1=1$ instead of $\beta_0<\beta_1$ based on
\begin{align}
-\frac{1}{2} \log (1+ JP\beta_1)< -\frac{1}{2} \log (1+ JP \beta_0).
\end{align}
This leads to a contradiction, and therefore we rule out the minimizers $\beta_0$, and the original problem becomes equivalent to the problem
\begin{align}
\min\limits_{\beta \in [1, J]: \ell(\beta,J,P) \leq 0 }\left\{ -\frac{1}{2} \log (1+ JP \beta) \right\}.
\end{align}  
\end{enumerate}
$\hfill \square$


\section*{Acknowledgments}
The authors wish to thank O.~L\'ev\^eque for helpful discussions and M. Wigger for pointing out an issue in an earlier version of the proof. This work was supported in part by the Swiss National Science Foundation under Grant 169294, Grant P2ELP2\_165137. G.~Kramer was supported in part by an Alexander von Humboldt Professorship endowed by the German Federal Ministry of Education and Research.


\bibliographystyle{IEEEtran}
\bibliography{IEEEabrv,nit_all}

\end{document}